\documentstyle[twoside,fleqn,espcrc2]{article}

\newcommand{\AmS}{{\protect\the\textfont2
  A\kern-.1667em\lower.5ex\hbox{M}\kern-.125emS}}

\title{Effects of an intermediate scale in SUSY grand unification }

\author{N. G. Deshpande$^{{\rm a}\, \, *}$,
                                                B. Dutta\address{Institute
of Theoretical Science, University of
                                                Oregon, Eugene, OR 97403,
USA \hfill {\bf OITS-606}}
        \thanks{Supported by DOE grant DE-FG06-854ER 40224.},
        and
        E. Keith\address{ Department of Physics, University of
                                                California, Riverside, CA
92521, USA \hfill {\bf UCRHEP-T167}}
                                        \thanks{Supported by DOE grant
DE-FG02-94ER 40837.}}

\begin{document}
%\preprint{\vbox{\hbox{OITS-602}\hbox{UCRHEP-T165}\hbox{hep-ph 9605386}}}
\begin{abstract}
We discuss the production of lepton flavor violation and  EDMs and the
viability of the
$b-\tau$ unification hypothesis in SUSY grand unification with an
intermediate gauge symmetry
breaking scale.
\end{abstract}

\maketitle

Supersymmetric grand unified theories (GUTs) with intermediate gauge
symmetry breaking scales are attractive because they resolve a few longstanding
problems and possess some desirable phenomenological features. For example, in
models where the intermediate breaking scale
$M_I \sim 10^{10}-10^{12}$ GeV, one can naturally  get a neutrino mass in the
interesting range of
$\sim 3-10$ eV, which could serve as hot dark matter to explain the observed
large scale structure formation of the universe. In models without an
intermediate gauge symmetry, in principle one could  produce a tau-neutrino
Majorana mass that is much less than the GUT breaking scale as for example via
non-renormalizable operators involving Higgs in the SO(10) ${\overline {16}
}$
representation or a  small and carefully chosen  Yukawa coupling to a
${\overline {126}}$ field. However,  this would suffer from the further
problem of
abandoning
$b-\tau$ Yukawa coupling unification except possibly in the case of high
$\tan\beta$ \cite{[ML10]}. The window
$\sim 10^{10}-10^{12}$ GeV is also of the right size for a hypothetical
PQ-symmetry to be broken so as to solve the strong CP problem without creating
phenomenological or cosmological problems. Models which allow
even lower intermediate gauge symmetry breaking scales \cite{[DKR],[EMa]}
e.g. $M_I\sim 1$ TeV
are also interesting since they predict relatively light new gauge fields,
as for
example SU(2)$_R$ charged gauge bosons
$W_R$. A further motivation for studying
GUTs with intermediate gauge symmetry breaking scales is that some
scenarios may allow
lower values of $\alpha_s$ as preferred by some experiments.

In this talk based on  Ref. \cite{[1],[2],[3]}, we will discuss
lepton flavor violation, the production of an electron and neutron electrec
dipole moment
(EDM), and the viability of the GUT scale  $b-\tau$ unification hypothesis
in GUTs
in
 which the $T_{3R}$ and $T_{B-L}$ generators are unbroken  above $M_I$. We
will always assume
that supersymmetry is broken via soft breaking terms introduced at a super
high scale. We
shall assume that the    soft breaking terms at the high scale at which
they are introduced
are flavor blind  and  CP invariant. As examples, in Ref.
\cite{[1],[2],[3]} we have used four different models with intermediate
gauge symmetry
breaking scales.

It has recently been pointed out \cite{[LJdhAS]} that significant lepton
flavor violation, as well a electron and neutron EDMs, can arise in
supersymmetric (SUSY) grand
unified theories. The origin of this flavor violation resides in the
largeness of the top
Yukawa coupling and the assumption that supersymmetry is broken by flavor
uniform soft
breaking terms communicated to the visible sector by gravity at a scale $M_X$.
Assuming that $M_X$ is the reduced Planck scale which is much greater than
$M_G$,
renormalization effects cause the third generation multiplet of squarks and
sleptons which
belong to the same multiplet as the top in the grand unified theory (GUT)
to become lighter
than those of the first two generations. The slepton and the charged lepton
mass matrices can
no longer be simultaneously diagonalized thus inducing lepton flavor violation
through a suppression of the GIM mechanism in the slepton sector. This effect is
more pronounced in SO(10) models than in SU(5) where the left-handed slepton
mass matrices remain degenerate. The evolution of soft terms from $M_X$ to $M_G$
causes these flavor violations, which disappear when $M_X=M_G$. Here,
we explore another class of theories which are SUSY SO(10) GUTs which break down
to an intermediate gauge group $G_I$ before being broken to the Standard Model
(SM) gauge group at the scale $M_I$. In this class of theories, even if
$M_X=M_G$, lepton flavor violation arises due to the effect of the third
generation neutrino Yukawa coupling on the evolution of the soft leptonic terms
from the grand unification scale to the intermediate scale. Depending on the
location of the intermediate scale
$M_I$ and the size of the top Yukawa coupling at $M_G$, these rates can be
within one order of magnitude of the current experimental limit. Our results
 also indicate that if
$M_X>M_G$ in SUSY SO(10) models with an intermediate scale, the predicted rates
of lepton violating processes are further enhanced. We will concentrate on the
decay
$\mu\rightarrow e\gamma$ as an example since experimentally it is likely to be
the most viable.

With $G_I$=SU(2)$_L\times$SU(2)$_R\times$SU(4)$_C$ ($\{ 2_L\, 2_R\,
4_C\}$), the quarks
and leptons are unified. Hence, the $\tau$-neutrino Yukawa coupling is the same
as the top Yukawa coupling. Through the renormalization group equations (RGEs),
the effect of the large
$\tau$-neutrino Yukawa coupling is to make the third generation sleptons lighter
than the first two generations, thus mitigating the GIM cancellation in one-loop
leptonic flavor changing processes involving virtual sleptons. Although the
quarks and leptons are not unified beneath the GUT scale when
$G_I$=SU(2)$_L\times$SU(2)$_R\times$U(1)$_{B-L}\times$SU(3)$_c$ ($\{ 2_L\, 2_R\,
1_{B-L}\, 3_c\}$), the same effect is produced from the assumption that the
top quark
 Yukawa coupling is equal to the $\tau$-neutrino Yukawa coupling at the GUT
scale. In Ref. \cite{[1],[2]}, one can see that such models can easily
predict rates of
lepton flavor violation that are within an order of magnitude beneath
experimtal limits and
can sometimes even put limits on the allowable parameter space.

When we calculate the EDM of the electron the  above stated  principle applies,
but we must also consider the phases at the  gaugino-slepton-lepton vertices.
Likewise, to generate the EDM for the neutron one needs the third
generation down
squark to be lighter than those of the other two generations, which occurs
due to
the large top Yukawa coupling, and  new phases at the gaugino-squark-quark
vertices. In fact, whenever there is  an intermediate scale, irrespective
of $G_I$   such
phases  are generated. The reason for this is that right-handed quarks or
leptons are unified
in a multiplet in a given generation. The superpotential for an
intermediate gauge symmetry
breaking model can be written (when $G_I=\{ 2_L\, 2_R\, 4_C\}$) as
$W_Y={\bf \lambda_{F_u}}{
F}{ \Phi_2}{ {\bar F}} +
 {\bf\lambda_{F_d}}{ F}{ \Phi_1}{{\bar F}}$,
where $F$ and ${\bar F}$ are the superfields containing the
standard model fermion fields and transform as $(2,1,4)$ and $(1,2,{\bar 4})$
respectively and we have suppressed the generation and gauge group indices. We
choose to work in a basis where
${\bf
\lambda_{F_u}}$ is diagonal in which $W_Y$ can be expressed as $W_Y={ F}{\bf
{\bar\lambda_{F_u}}}{ {\bar F}}{
\Phi_2} +
 { F}{\bf {\bf U^*}{\bar\lambda_{F_d}}}{\bf U^{\dag}}{{\bar F}}{ \Phi_1}$.
The matrix {\bf U}
is a general
$3\times3$ unitary matrix with 3 angles and 6 phases. It can be written as
${\bf U}={\bf S^{\prime *}} {\bf V} {\bf S}$,  where {\bf V} is the CKM
matrix and ${\bf S}$
and
${\bf S^\prime}$ are  diagonal phase matrices. At the weak scale,
 we have the Yukawa terms:
\begin{eqnarray} W_{\rm MSSM}&=&Q{\bf \bar\lambda_u}{\bf U^c}H_2 +Q{\bf V^*}{\bf
\bar\lambda_d}{\bf S}^2{\bf V^{\dag}}D^cH_1\nonumber\\ &+&E^c{\bf V_I^*}{\bf
\bar\lambda_L}{\bf S}^2{\bf V_I^{\dag}}LH_1\, ,
 \end{eqnarray} where the ability to reduce the number of phases by redefinition
of fields has been taken advantage of to the fullest extent possible,
 ${\bf S}^2\equiv {\rm Diag.}\left(
               {\rm e}^{i\phi_d}, {\rm e}^{i\phi_s}, 1   \right)$
 is a diagonal phase matrix with two independent phases, and ${\bf V_I}$ is the
CKM matrix at the scale $M_I$. It is not possible
to do a superfield rotation on
$D^c$ or
$L$ to remove the right handed angle since at $M_I$ the third diagonal element
of the scalar mass matrices
${\bf m_D^2}$ and
${\bf m_L^2}$ develop  differently from the other two diagonal elements due to
large top Yukawa coupling RGE effects. When
$G_I=\{ 2_L\, 2_R\,
1_{B-L}\, 3_c\}$,  the additional CKM-like
phases will be generated in exactly the same way  as described above.
Results for specific
examples can be found in Ref. \cite{[3]}, where one can see that highly
significant EDMs can be
produced by the physics of the intermediate scale  gauge symmetry.

Now we discuss  the viability of $b-\tau$ unification hypothesis. In this
talk, we will
concentrate on the following two  models: (1)  the model of Ref. \cite{[2]}
which has
$G_I$=SU(2)$_L\times$SU(2)$_R\times$SU(4)$_C$ with
$M_G\approx M_{\rm string}$ and
$M_I\sim 10^{12}$ GeV, and (2) the model of  Ref. \cite{[DKR]} which has
$G_I$=SU(2)$_L\times$SU(2)$_R\times$U(1)$_{B-L}\times$SU(3)$_C$ and allows
$M_I$ to have any value between the TeV scale and $M_G$.   The value for
$m_b^{\rm pole}$ from the existing data is
$m_b^{\rm pole}$=4.75$\pm$ .05, and we use
$m_\tau =1.777$ GeV. The predicted $m_b$ mass in these intermediate scale models
mainly depend on 3 factors: the value of
$\lambda_{t_{G}}$,
$\alpha_s (M_Z)$ and the location of the intermediate scale $M_I$. Using larger
values of
$\lambda_{t_{G}}$ of course lowers the $m_b$ mass, while using larger values of
$\alpha_s$ increases it. For these models at the scale
$M_G$,  we  use the maximum perturbative value for the top Yukawa coupling
which is about  3.54. For  the model of Ref. \cite{[2]}, since leptons and
down quarks are
unified in the same multiplet at the intermediate scales we have
$\lambda_b=\lambda_\tau\neq \lambda_t=\lambda_{\nu_{\tau}}$ for the low
$\tan\beta$ version of that model. We find $m_b^{\rm pole}=4.78$ GeV. For
the large
$\tan\beta$ version, we have
$\lambda_t=\lambda_b=\lambda_\tau=\lambda_{\nu_{\tau}}$ instead, and find
$m_b^{\rm pole}=4.80$ GeV. We find that the model of Ref. \cite{[2]} is
able to provide
very reasonable b-quark mass predictions since $\alpha_s$ is of  moderate
values and
since down to the scale $M_I$ the relation $\lambda_b =\lambda_\tau$ exists
intact.    For the model of Ref.
\cite{[DKR]} with
$\lambda_t=\lambda_b=\lambda_\tau=\lambda_{\nu_{\tau}}$ at $M_G$, in Fig.
2(a) of Ref.
\cite{[3]} we   plot the
$m_b^{\rm pole}$ mass as a function of
$M_I$ since the intermediate gauge symmetry breaking  scale in that model can
lie anywhere between the weak scale and the GUT scale. We find that that
the b-quark
mass at first increases as the intermediate gauge symmetry breaking scale  moves
away from the GUT scale. But, it then reaches a peak value when the intermediate
scale is about
$10^8$ GeV, and then for  $M_I$ less than that scale it decreases. The reason
for this behavior can be found in the RGEs for  $\lambda_b$ and
$\lambda_\tau$. The RGE for $\lambda_b$ feels the influence of the large top
Yukawa coupling while
$\lambda_\tau$ instead feels the influence of the
$\tau$ neutrino coupling. Though the magnitude of the top and the $\tau$
neutrino couplings are same at the GUT scale, the $\tau$-neutrino coupling
decreases faster than the top Yukawa coupling and reaches its fixed point
sooner. If the Intermediate breaking scale is decreased  $\lambda_b (M_I)$ would
also decrease, however  $\lambda_\tau (M_I)$ would not decrease as much. So,
effectively the mass of $m_b$ decreases, since
$m_b$ mass depends on the ratio of $\lambda_b$ to $\lambda_\tau$. We further
note that the interesting values for the intermediate gauge symmetry breaking
scale
$M_I\sim 1$ TeV and
$M_I\sim 10^{12}$ GeV can both give good values for the b-quark mass. Effects of
this low intermediate scale could be observed in the future colliders. In Fig.
2(b) of Ref. \cite{[3]}, we assume the possibility of the model of Ref.
\cite{[DKR]} allowing a
range of values for
$\tan\beta$ in order to plot the $m_b^{\rm pole}$ as a function of
$\tan{\beta}$ for the interesting case of $M_I=10^{12}$ GeV. We see that larger
values of $\tan\beta$ are preferred and give very reasonable values for the
b-mass. Of course as in Ref. \cite{[BM]}, one could purposefully
construct models with
$G_I$=SU(2)$_L\times$SU(2)$_R\times$U(1)$_{B-L}\times$SU(3)$_C$  that
have $M_I\sim 10^{12}$ GeV and lower values for
$\alpha_s$ so as to improve the b-quark mass prediction for low values of
$\tan\beta$.

\end{document}